\documentclass[reprint,amsmath,amssymb,aps,pra]{revtex4-2}
\usepackage{graphicx}% Include figure files
\usepackage{dcolumn}% Align table columns on decimal point
\usepackage{bm}% bold math
\usepackage{amsmath}
\usepackage{amsfonts,amssymb} 
\usepackage{tabularx}
\usepackage{booktabs}
\usepackage{multirow}
\usepackage{color}
\usepackage{siunitx}
\usepackage{tabu}
\usepackage[ruled]{algorithm2e}
\usepackage{subfigure}
\usepackage[T1]{fontenc}
\usepackage{color}
\usepackage{hyperref}
\usepackage{braket}

\begin{document}

\title{Variational Quantum Circuits Enhanced Generative Adversarial Network}

\author{Runqiu  Shu$^{1,2}$}
\author{Xusheng Xu$^{2}$}
\author{Man-Hong Yung$^{2,3}$}
\email{yung@sustech.edu.cn}
\author{Wei Cui$^1$}
\email{aucuiwei@scut.edu.cn}

\address{$^1$School of Automation Science and Engineering, South China University of Technology, Guangzhou 510641, China}
\address{$^2$Central Research Institute, 2012 Labs, Huawei Technologies}
\address{$^3$Shenzhen Institute for Quantum Science and Engineering}

\begin{abstract}
	Generative adversarial network (GAN) is one of the widely-adopted machine-learning frameworks for a wide range of applications such as generating high-quality images, video, and audio contents. However, training a GAN could become computationally expensive for large neural networks. In this work, we propose a hybrid quantum-classical architecture for improving GAN (denoted as QC-GAN). The performance was examed numerically by benchmarking with a classical GAN using MindSpore Quantum on the task of hand-written image generation. The generator of the QC-GAN consists of a quantum variational circuit together with a one-layer neural network, and the discriminator consists of a traditional neural network. Leveraging the entangling and expressive power of quantum circuits, our hybrid architecture achieved better performance (Frechet Inception Distance) than the classical GAN,  with much fewer training parameters and number of iterations for convergence. We have also demonstrated the superiority of QC-GAN over an alternative quantum GAN, namely pathGAN, which could hardly generate 16$\times$16 or larger images. This work demonstrates the value of combining ideas from quantum computing with machine learning for both areas of Quantum-for-AI and AI-for-Quantum.
\end{abstract}

\maketitle

\section{Introduction}
Machine learning has made great achievements in various fields recently. However, the number of training parameters, the scale of the model, and the training time consumed lead to a significant challenge for computing power. This leads to the necessity of searching for more computing power. Quantum computing introducing quantum property may solve problems with a super-polynomial speedup relative to classical computers\cite{preskill2012quantum}. Besides, quantum systems produce typical patterns that classical systems are thought not to produce efficiently~\cite{biamonte2017quantum} which may provide new advantages on machine learning tasks. Therefore, there is a combination of machine learning and quantum, called quantum machine learning(QML)~\cite{biamonte2017quantum,lamata2020quantum}.
QML has the potential of accelerating data analysis~\cite{cerezo2022challenges}, especially for quantum data, which has spurred much attention in researching corresponding algorithms. A lot of QML algorithms have been proposed recently, including quantum convolution neural network~\cite{cong2019quantum}, quantum classifier~\cite{blank2020quantum}, and quantum generative adversarial network (QGAN)~\cite{lloyd2018quantum,dallaire2018quantum}, among others.

Generative adversarial network (GAN) is a well-known machine learning model proposed by Goodfellow~\cite{goodfellow2014generative}. It has outstanding performance in various challenging tasks, especially in image generation and video generation tasks. However, the demand of GAN for model expressiveness leads to a dramatic increase in model scale and training parameters, which raises the difficulty of model training leading to task failure. QGAN, as a quantum version of GAN, has more advantages in sampling and generating discrete data than its classical counterpart. In addition, it has been proved the potential exponential speedup theoretically~\cite{lloyd2018quantum}. These lead to extensive research on QGAN. For example, the learning of the classical or quantum data~\cite{benedetti2019adversarial, zeng2019learning, zoufal2019quantum,ahmed2021quantum}, using the entangling power of quantum circuits to overcome issues of non-convexity and mode collapse~\cite{niu2022entangling}, discovering small molecular drugs~\cite{li2021quantum}, data enhancement~\cite{nakaji2021quantum} and anomaly detection with QGAN~\cite{herr2021anomaly}. Previous works have demonstrated the unique advantages and wide applications of the QGAN algorithm in handling both classical and quantum tasks. Moreover, the successful experimental implementation in superconducting system~\cite{huang2021quantum, hu2019quantum} has demonstrated the feasibility of the algorithm. However, we are entering the Noisy Intermediate-Scale Quantum (NISQ)~\cite{preskill2018quantum} era, which means a limited number of qubits and quantum gates, and a noisy environment. It is difficult to tackle complex real-world tasks with pure quantum circuits QGAN. For the image generation task, Ref.~\cite{huang2021experimental} proposed the idea of using multiple quantum sub-generators. Each sub-generator is assigned to a part of the image generation, which is finally connected to a complete image. Ref.~\cite{stein2021qugan} reduced the dimensionality of the image by principal component feature extraction before entering the quantum circuits, which reduces the requirement for quantum resources. Both of them inevitably lead to the loss of information thus reducing the clarity of the generated images. In addition, the size of the generated images is limited. These impair the excellent performance of GAN on image generation tasks. Therefore, exploring the utilization of quantum computing advantages to enhance the performance of machine learning algorithms under limited quantum resources becomes an important issue.

In this paper, we explore the enhancement of classical algorithms by quantum computing. More specifically, we develop a hybrid quantum-classical architecture(QC-GAN) which makes improvements to the classical GAN on image generation. QC-GAN also consists of a generator and a discriminator. Differently, the generator introduces variational quantum circuits~\cite{benedetti2019parameterized} to exploit quantum computation. We successfully implement the generation experiments on a handwritten dataset. The experimental results show that benefitting from quantum computational power this architecture generates better quality images with much fewer training parameters and iterations than classical GAN. Moreover, in comparison with other structures of QGANs, such a combination obtains better image generation results.

The rest of the paper is organized as follows: In Section 2, we briefly introduce the theory of generative models, GAN, and variational quantum circuits. We describe the details of the proposed QC-GAN in section 3. In Section 4 we demonstrate the numerical experiments and results. Finally, the conclusion is presented in Section 5. 

\section{PRELIMINARY}
\subsection{Generative models}
The generative models aim to obtain data features and generate distributions similar to the sample. Considering sampling ${m}$ samples ${X=\{x_1,\dots,x_m\}}$ from the distribution ${p_{data}}$, generative models attempt to generate a distribution ${p_{model}}$ that is similar to the input distribution ${p_{data}}$. Most generative models work via the principle of maximum likelihood. For a dataset ${X}$ containing ${m}$ training examples, the likelihood is written as
\begin{equation}
  \prod_{i=1}^m p_{model}\left(x_i, \theta\right).\label{eq1}
\end{equation}
To learn the target distribution, generative models optimize parameters by maximizing the likelihood of the training data. The object function of parameters is written as follows, 
\begin{equation}
  \begin{split}
  \boldsymbol{\theta}^*&=\underset{\boldsymbol{\theta}}{\arg \max } \prod_{i=1}^m p_{\text {model }}\left(\boldsymbol{x}^{(i)} ; \boldsymbol{\theta}\right)\\
  &=\underset{\boldsymbol{\theta}}{\arg \max } \log \prod_{i=1}^m p_{\text {model }}\left(\boldsymbol{x}^{(i)} ; \boldsymbol{\theta}\right)\\
  &=\underset{\boldsymbol{\theta}}{\arg \max } \sum_{i=1}^m \log p_{\text {model }}\left(\boldsymbol{x}^{(i)} ; \boldsymbol{\theta}\right).\label{eq2}
  \end{split}
\end{equation}
For convenience, the original form is converted into log space, where we have a sum rather than a product.
Finding ${\theta^*}$, the generative model enables to approximate ${p_{data}}$ with ${p_{model}}$.

\subsection{Classical generative adversarial networks }
According to whether an explicit form of the  density function is used or not, generative models can be classified as explicit and implicit. GAN belongs to the implicit density model. Compared with other generative models, such as variational autoencoder~\cite{kingma2013auto} and diffusion model~\cite{ho2020denoising}, GAN uses the learning ability of neural networks to provide the target distribution. It avoids the use of Markov chains and makes the calculation more efficient. In addition, it generates visually clearer samples. Therefore, it has a wide range of applications, including image and video generation~\cite{duvenaud2015advances},
super-resolution sensing~\cite{wang2018esrgan}, object recognition~\cite{bai2018sod}, 
anomaly detection~\cite{schlegl2017unsupervised} and so on.

It is difficult to compute the generated probability distribution using the maximum likelihood estimation described above, so GAN introduces a discriminator. Therefore, the general GAN consists of a generator ${G}$ and a discriminator ${D}$. It accomplishes generation tasks through an adversarial game of ${G}$ and ${D}$. During the training, ${G}$ is trained to maximize the probability that ${D}$ misclassifies the generated data as the real data. 
Correspondingly, ${D}$ is trained to maximize the probability of successfully classifying real data, while minimizing the probability of misclassifying generated data. The optimization of GAN can be summarized as
\begin{equation}
	\begin{split}
	  \mathop {\min }\limits_G \mathop {\max }\limits_D V(D,G) = {\rm{ }}{{\rm{E}}_{x\sim{p_r}(x)}}[\log D(x)] \\
	  + {{\rm{E}}_{z\sim{p_z}(z)}}[\log (1 - D(G(z)))]. \label{eq3}
	\end{split}
  \end{equation}
Using the noise ${z}$ sampled from the prior distribution ${{p_z}(z)}$ as input, ${G}$ generates the probability distribution and attempts to make it close to the real probability distribution ${p_r}$. Meanwhile, ${D}$ attempts to distinguish the real probability distribution ${p_r}$ and the generated probability distribution. By training the generator and discriminator alternately, it will end up at a Nash equilibrium point theoretically. The discriminator has a $1/2$ probability of discriminating correctly for the probability distribution generated by the generator.

\subsection{Variational quantum circuits}
\begin{figure}[htbp]
  \centering
  \includegraphics[scale=0.3]{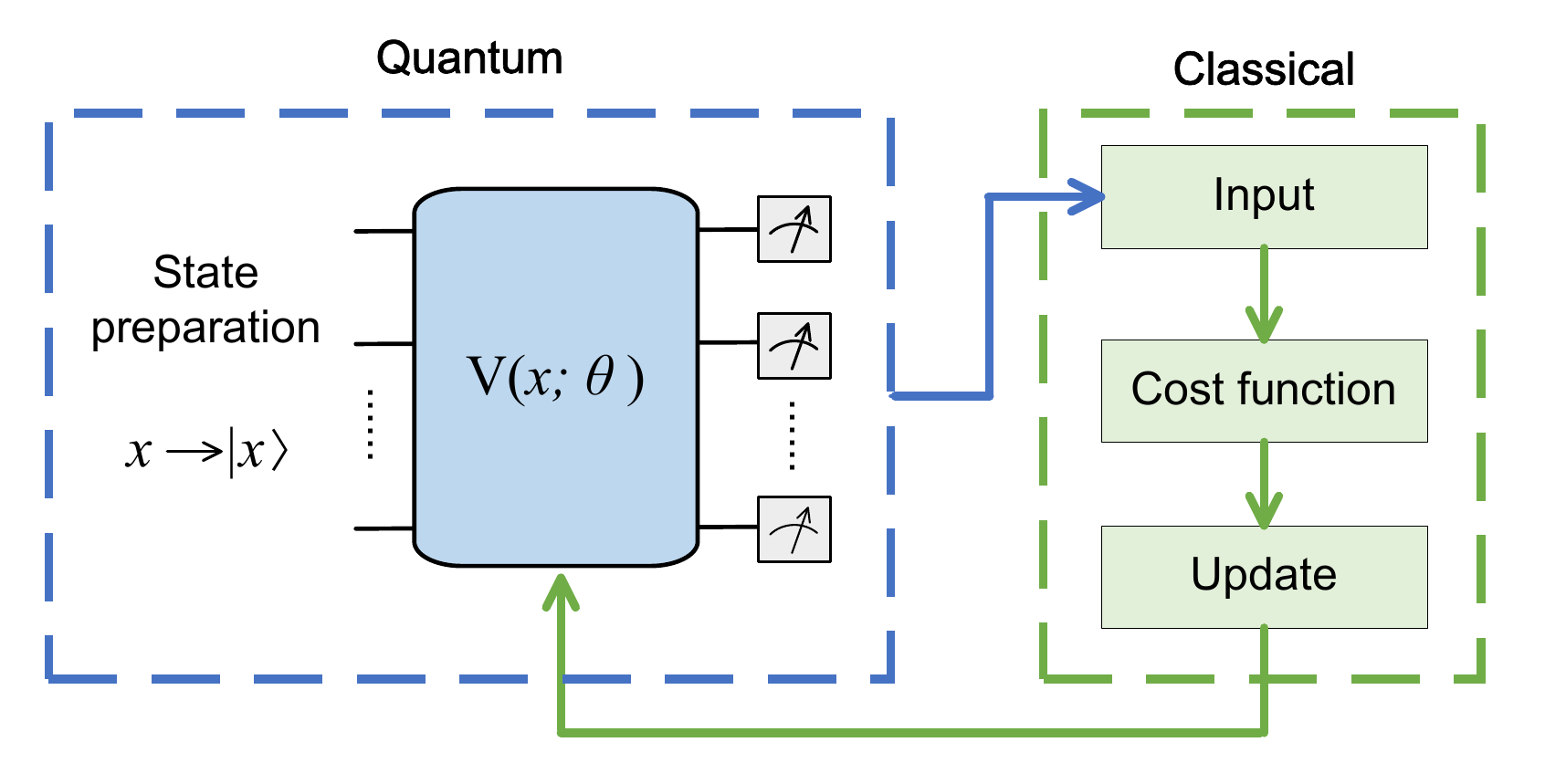}
  \caption{\label{fig1} The algorithms based on PQCs are optimized in classical and quantum systems.}
\end{figure}

Variational quantum circuits, also known as parameterized quantum circuits (PQCs), are typically composed of some gates with free parameters. The gates of circuits are arranged in a particular combination according to the task requirements and quantum resources. The parameter optimization process involves both classical and quantum systems, see Figure~\ref{fig1}. Firstly, the initial quantum state ${\ket{x}}$ is prepared by encoding classical data. Then parameterized circuits ${V(\theta)}$ act on the initial state ${\ket{x}}$. To obtain the results, it is necessary to select an appropriate observable operator to measure the output. Finally, the model parameters will be optimized according to the cost function and the updates of parameters are implemented by classical computers.

In the case of limited quantum resources, the hybrid approach based on PQCs has been proven even at low depths, the circuits are capably expressive. Thus it has been applied to a variety of problems and has been successful in chemistry, combinatorial optimization,  and machine learning. 

\begin{figure*}[ht]
  \centering
  \includegraphics[scale=0.28]{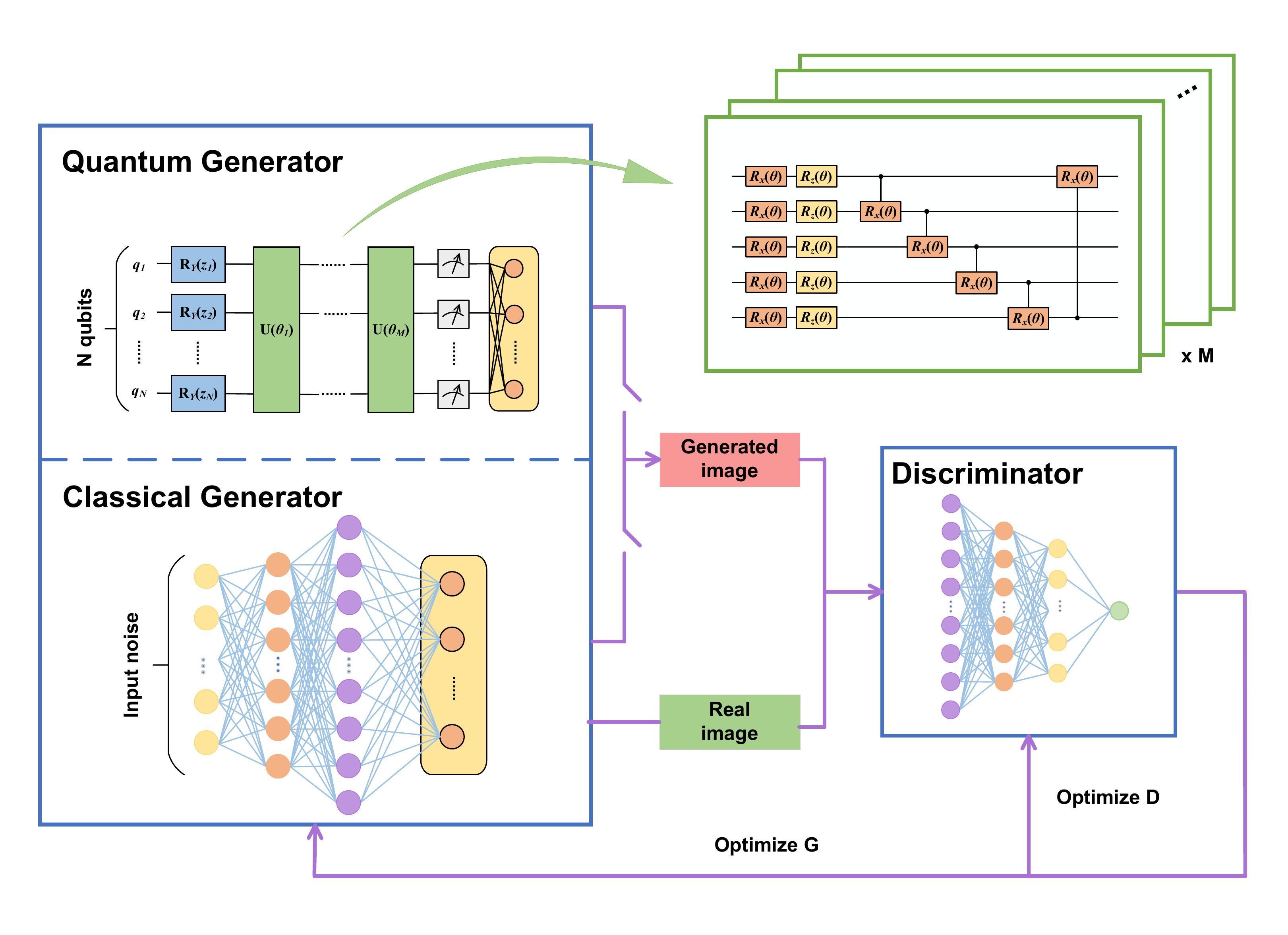}
  \caption{\label{fig2} The overall architecture and training process of QC-GAN and the the classical GAN used in the experiment.}
\end{figure*} 

\section{QUANTUM GENERATIVE ADVERSARIAL NETWORK}
\subsection{The architecture of the model}
Similar to the classical GAN, QGAN consists of 
two components: generator ${G}$ and discriminator ${D}$. 
Differently, the quantum circuit replaces ${G}$ or 
both of ${G}$ and ${D}$.
Considering the generation of handwritten digital images, 
we propose a quantum-classical hybrid architecture.
 It consists of a generator containing a quantum variational circuit with a layer of neural network and a classical 
discriminator, as shown in Figure~\ref{fig2}. 
The process of generation training consists of three parts: 
encoding, variational evolution and measurement, and post-processing.

\subsubsection{Encoding}
In QGAN algorithm that process a classical input ${z{\sim}p_z(z)}$, 
it has to be encoded to a quantum state at first.
This process is considered as a nonlinear mapping from a 
low-dimensional to a high-dimensional Hilbert space.
Thus, a small size noise input can be mapped to a more expressive space.
There are many kinds of encoding methods, and the amplitude 
encoding method is a common strategy. 
It uses the probability amplitudes of the quantum state 
to encode the classical input data. 
A quantum state in ${N}$ qubits can encodes 
classical data of ${2^N}$ size. 
However, this method requires to operate quantum circuit with
exponential depth, which might destroy the quantum advantage
~\cite{plesch2011quantum, nakaji2022approximate}.

Considering that the input of QGAN algorithm is a random noise vector, it has a lower requirement for accuracy.  
It is one of the advantages of QGAN compared to other quantum machine learning algorithms.
We simply encode the input into a quantum circuit as angle parameters.
This implies that only one layer of depth is required, easing the bottleneck
of data input.
For the input ${z=\{z_1,z_2,\dots,z_N\}}$, it is encoded as a variable 
in the rotation operator ${R_y}$ as
\begin{equation}
  {\ket{\psi_z}} =\otimes_{i=1}^{N} R_{y}\left(z_{i}\right)\ket{0}.
  \end{equation}

\subsubsection{Variational evolution}
The evolution of the input quantum state ${\ket{\psi_z}}$
is performed on the quantum circuits. 
Most of quantum circuits consist of controlled 
rotate gates and entangled gates. 
The introduction of quantum entanglement properties
increases the correlation between information. 
It can produce patterns that can not be 
efficiently simulated by classical systems~\cite{liu2018differentiable}.

In addition, the different combinations of gates will 
produce different effects. An efficient quantum 
circuit can explore a wider target space with 
more expressive power. By comparing several different 
circuit combinations, we use the circuit block 
${U(\theta)}$ consists of ${R_x}$ and ${R_z}$. Repeating ${M}$ times 
consists of  the overall quantum circuit ${V(\theta)}$.
Applying the quantum circuit ${V(\theta)}$
on ${\ket{\psi_z}}$, we obtain the quantum state
\begin{equation}
  {{\ket{\psi_z,\theta}} =V(\theta){\ket{\psi_z}}}.
\end{equation}
The parameters ${\theta}$ in the circuit will be 
optimized during the training process, 
thus realizing the evolution of the initial 
state to the target state.

\subsubsection{Measurement and post-processing}
To obtain the classical expression of the probability distribution, the measurement process follows the variational evolution. The values are obtained by measuring on the selected observable operator. 
The evolution of quantum circuits is linear, which makes it difficult to generate several complex data distributions. Moreover, due to the limited quantum resources, it is unrealistic to deal with large-scale data generation. The second part of our generator is neural networks, which can be considered as a post-processing of the output of the quantum circuits. The post-processing performs a mapping and a nonlinear change of the quantum circuits output to the data dimension. Such processing makes the best available use of quantum circuits while overcoming the lack of resources.

\subsection{The training of the model}
Classical random variables are encoded into quantum 
states and then evolve in the quantum circuit. 
Recently available quantum devices have 
severe restrictions, including a limited number of qubits 
and noisy processes that limit the depth of the circuit. 
The optimization of quantum circuits is usually performed 
using classical optimizers. 
The parameters of the quantum circuits are optimized 
according to the designed cost function. 
The generator needs to generate a probability distribution as real as 
possible, and the discriminator needs to distinguish between the real 
distribution and the generated distribution as much as possible. 
Their cost functions are designed separately as Eq.(6) and Eq.(7). 
\begin{equation}
  {L_G} =  - {{\rm E}_{x\sim{p_f}(x)}}\left[ {\log D(x)} \right],
  \end{equation}
\begin{equation}
  {L_D} =  - {{\rm E}_{x\sim{p_r}(x)}}\left[ {\log D(x)} \right] - {{\rm E}_{x\sim{p_f}(x)}}\left[ {\log D(x)} \right].
  \end{equation}

The training of QGAN is similar to GAN that generator and discriminator are trained alternately. The gradient of the loss function is used to update parameters of the generator and discriminator to improve their performance in the adversarial game.

\section{NUMERICAL EXPERIMENTS}
This section presents numerical simulations performed on a handwritten dataset, alongside a detailed analysis of the experimental results. Furthermore, a comparative analysis is conducted between our model and classical GANs, as well as other existing QGANs.

\subsection{Metics}
To assess the quality of our model's generated images, a suitable metric is required. In this paper, we have introduced the Frechet Inception Distance (FID)~\cite{Seitzer2020FID,heusel2017gans} , which measures the similarity between the generated and real distributions by calculating the distance in feature space. The calculation of FID score is 
\begin{equation}
  FID(x,g) = ||{\mu _x} - {\mu _g}||_2^2 + Tr({\sigma _x} + 
  {\sigma _g} - 2{({\sigma _x}{\sigma _g})^{0.5}}),
  \end{equation}
where ${x}$ and ${g}$ are the real image and generated image, respectively, $\mu $ is the mean and $\sigma $ is the covariance matrix. 
The lower score of FID, the more real the generated image is. 

\subsection{Experimental dataset and settings}
To demonstrate the generation ability of QC-GAN, we evaluate it on the MNIST dataset~\cite{lecun1998gradient}, which consists of 10 classes of grayscale images of handwritten digits, each with a size of 28$\times$28 pixels. During the experiment, each category of data serves as a training set.

The quantum circuits are constructed and trained using MindQuantum, a general quantum computing framework for building quantum neural networks and supporting their training and inference. The overall architecture of the model has been described above. To optimize the utilization of quantum resources, we set the number of qubits ${N}$ to 5 and the depth of the variational quantum circuit ${M}$ to 4. For the discriminator ${D}$, we use a fully connected neural network with two hidden layers, consisting of 64 units in the first layer and 32 units in the second layer. We employ the SGD optimizer to optimize the parameters.

\subsection{Experimental results and analysis}
\begin{figure}
	\centering
	\includegraphics[scale=0.36]{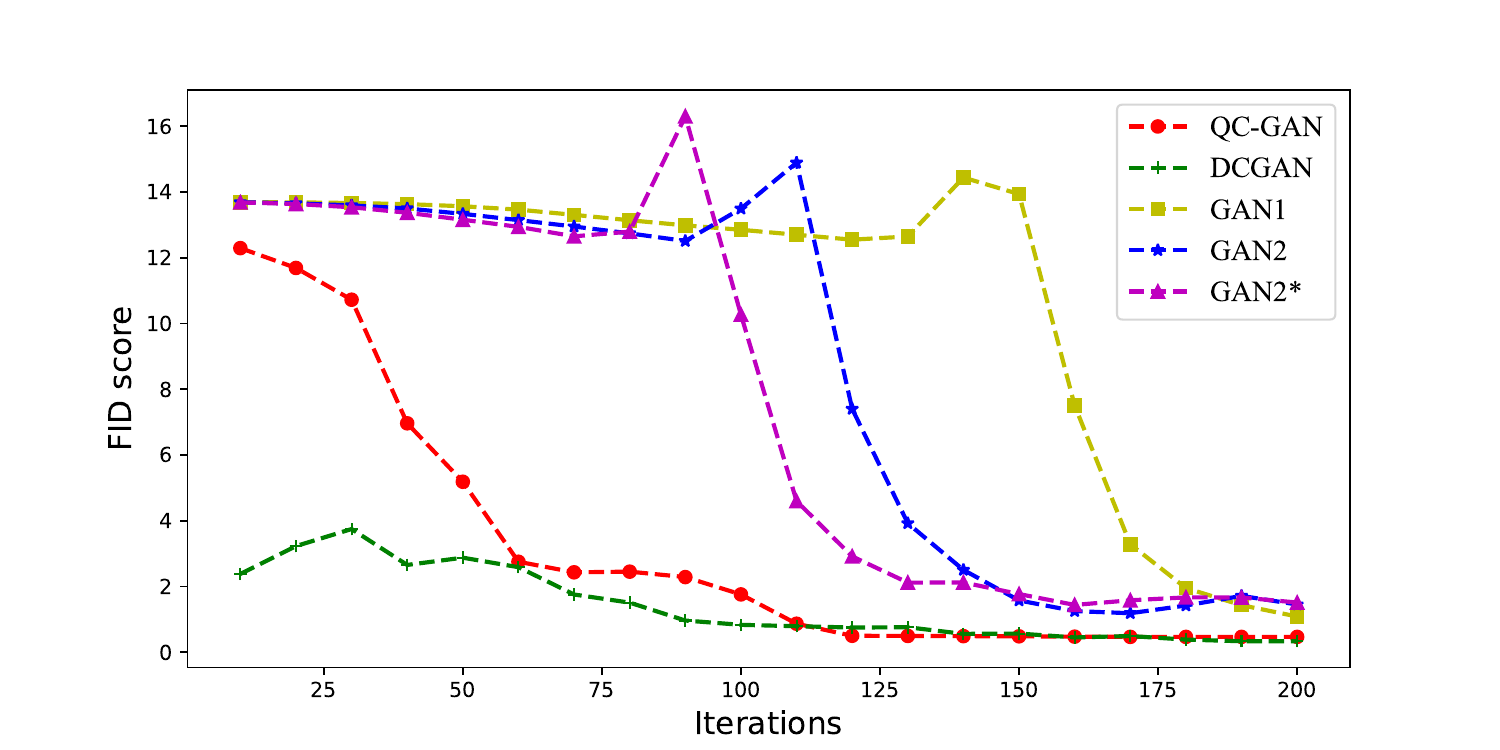}
	\caption{\label{fig3} The FID scores of the classical GAN with 80, 768 parameters excluding the last layer 
	and the QC-GAN with 65 quantum parameters. In addition, in the case of 768 parameters, the red line corresponds to an input 
	size of 100 and the blue line corresponds to an input size of 5.}
  \end{figure}

We compare QC-GAN with the classical GANs on the task of image generation. The structures of the compared generative models and the number of trainable parameter layers are shown in Table 1. The GAN1 and GAN2 corresponds to a general fully connected neural network, with different numbers of nodes in the hidden layers. DCGAN~\cite{radford2015unsupervised} is a convolutional neural network composed of multiple deconvolutional layers and batchnorm layers. Taking the generation of the digit 0 as an example, we use the FID score as a measure of image quality, and the experimental results are shown in Figure ~\ref{fig3}. The results show that QC-GAN has fewer iterations and better generated image quality compared to the general GANs. Compared with DCGAN, QC-GAN achieves a similar FID score with a shallower model depth and a significant reduction in trainable parameters.

\begin{table}[h!]
\begin{center}
	\caption{The layers with trainable parameters and the corresponding number of training parameters in various generative models. GAN2 and GAN2* correspond to different input sizes, other layers have the same structure.}
	\setlength{\tabcolsep}{5.5mm}
	\begin{tabular}{c|c|c}
	\toprule
	\textbf{Model} & \textbf{Layer} & \textbf{Parameters}\\
	\hline
	\multirow{2}{*}{QC-GAN} & QuantumLayer & 60\\
	&Linear&25872\\
	\hline
	\multirow{9}{*}{DCGAN} & Conv2 & 819200\\
	&BatchNorm&1024\\
	& Conv2 & 2097152\\
	&BatchNorm&512\\
	& Conv2 & 524288\\
	&BatchNorm&256\\
	& Conv2 & 131072\\
	&BatchNorm&128\\
	& Conv2 & 64\\
	\hline
	\multirow{3}{*}{GAN1} & Linear1 & 96\\
	&Linear2&1088\\
	&Linear3&50960\\
	\hline
	\multirow{3}{*}{GAN2} & Linear1 & 1536\\
	&Linear2&131584\\
	&Linear3&402192\\
	\hline
	\multirow{3}{*}{GAN2*} & Linear1 & 25856\\
	&Linear2&131584\\
	&Linear3&402192\\
	\bottomrule
	\end{tabular}
\end{center}
\end{table}  

In addition, we compare our QC-GAN with two other QGANs for the task of image generation. The patchGAN~\cite{huang2021experimental} uses a quantum generator while the discriminator is classical, and employs multiple quantum sub-generators to generate images. This approach alleviates some of the limitations imposed by the lack of quantum resources. However, each quantum sub-generator requires an auxiliary qubit for a nonlinear map, and the total number of qubits required for generating images with n pixel points is given by ${g(\log_{2}(n/g)+1)}$, where ${g}$ is the number of sub-generators. As for QuGAN~\cite{stein2021qugan}, both the generator and discriminator are quantum. It uses principal component analysis to extract features from the input samples, where each qubit represents a feature of the sample. For a p-dimensional feature vector, the number of qubits required is ${2p}$, which means that the required quantum resources are doubled. However, the generated images by QuGAN are often unclear due to the drastic dimensionality reduction of the images that compromises a large amount of information.

In our simulation experiments, we use pathGAN and QC-GAN to generate handwritten digits 0 and 1, respectively, and compare the generated images. When the size of the generated images is 8x8, pathGAN requires 4 sub-generators with 5 qubits per generator. When the size of the generated images increases to 16$\times$16, the number of qubits per sub-generator remains the same, and 16 sub-generators are used. For both experiments of generating images with the size of 8x8 and 16x16, we conduct simulations based on the model and parameter settings introduced earlier.

\begin{figure}
  \centering
  \includegraphics[scale=0.25]{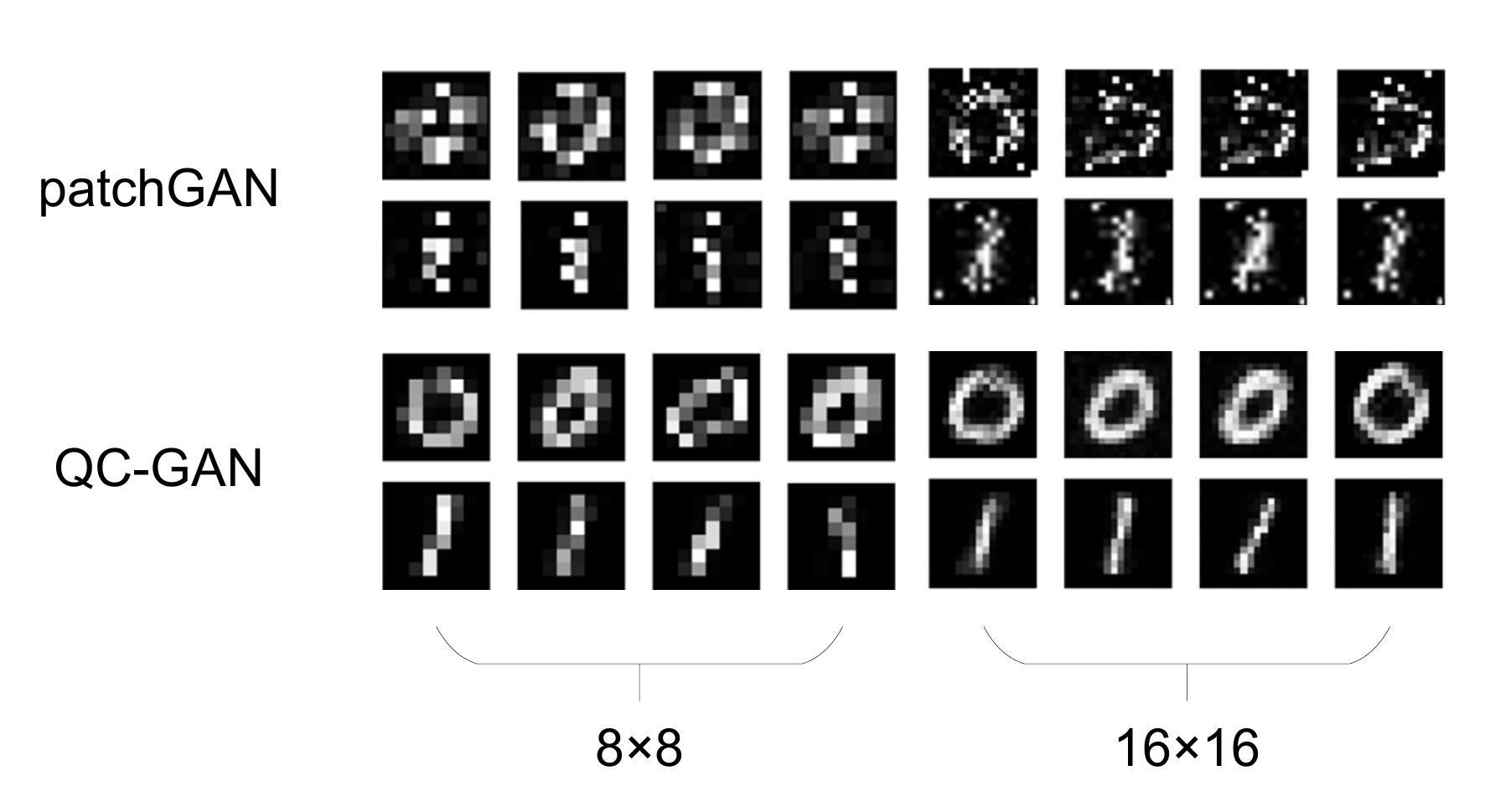}
  \caption{\label{fig4} The comparison of patchGAN and QC-GAN for image generation, including the generation of handwritten digit 0 and 1 of size 
  8$\times$8 and 16$\times$16.}
\end{figure}

As depicted in Figure~\ref{fig4}, the experimental results reveal that pathGAN and QC-GAN perform comparably for low-resolution image generation. However, as the image resolution increases, pathGAN requires a larger number of quantum resources and yields suboptimal outcomes in the generation task, while QC-GAN continues to generate high-quality images with clarity and fidelity.

It is important to note that the comparison is specific to the task of image generation and the performance of different QGANs may vary for other tasks. Additionally, the quantum resources required for a particular QGAN may depend on the specific model architecture and the size of the input data.

\begin{figure}
	\centering
	\subfigure[\label{5a}]{
		\includegraphics[scale=0.45]{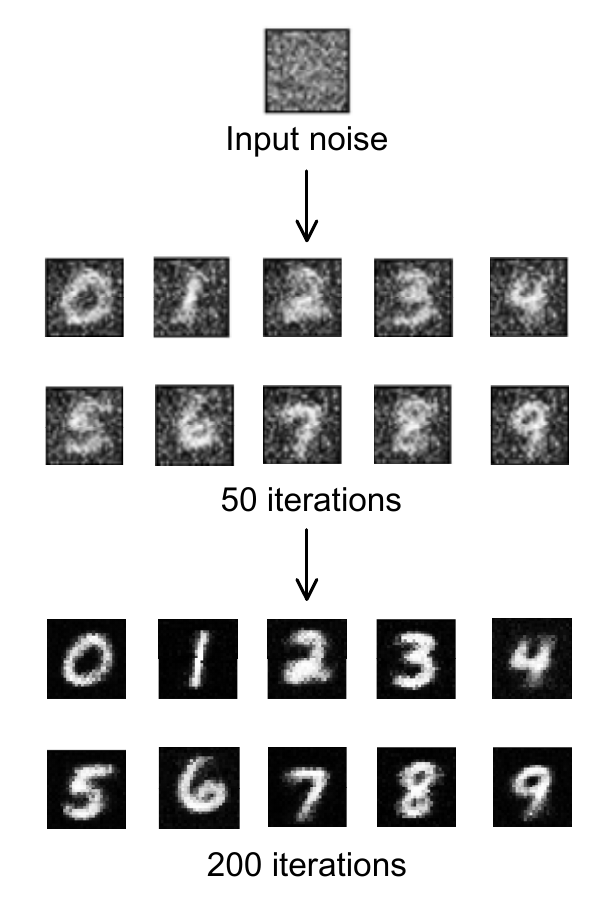}}
	\\
	\subfigure[\label{5b}]{
		\includegraphics[scale=0.36]{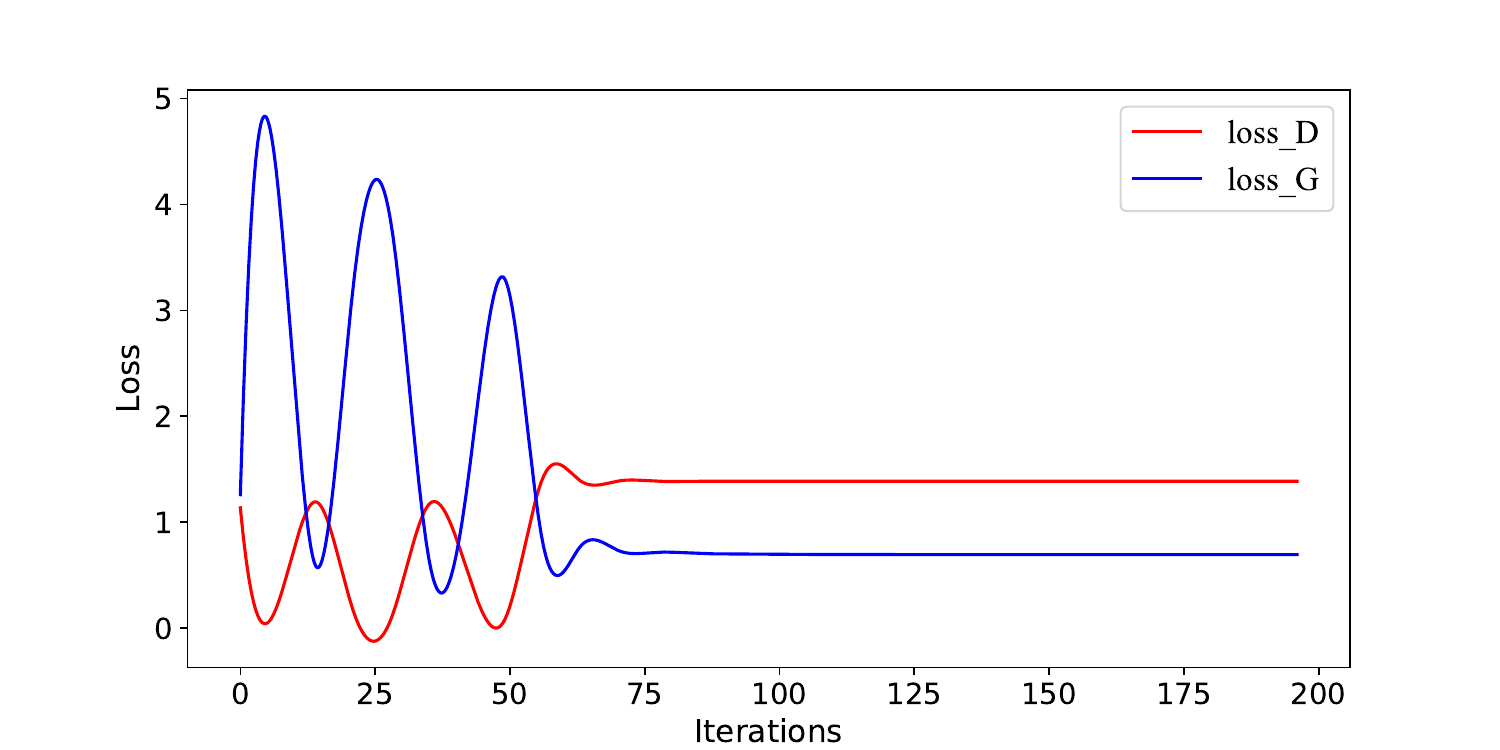} }
	\caption{\label{fig5}Handwritten digit 
	generation of 28$\times$28 size using QC-GAN. (a)The noise input gradually changes to clear digits as the number of iterations increases. (b)The change of generator and discriminator loss curves during the training process.} 
\end{figure}

Figure~\ref{fig5} illustrates the image generation process and the variation of the loss curve when using QC-GAN to generate handwritten digits of size 28x$\times$28. As the number of iterations increases, the noise distribution ${z{\sim}p_z(z)}$ gradually transforms into a clear image, even when using only 5 qubits. However, it is worth noting that more training iterations do not necessarily lead to better results.

Observing the training process of the model, we can discern the adversarial relationship between the generator ${G}$ and discriminator ${D}$: when one is improving, the other is declining, and vice versa. The loss curves of both ${G}$ and ${D}$ exhibit dramatic changes at the beginning of training, reflecting the training instability that is a common challenge for GANs and QGANs alike. Fortunately, careful selection of hyperparameters or incorporation of appropriate training tricks can alleviate this instability. After several rounds of significant changes, the loss curves eventually reach an equilibrium state.

By comparing with classical GANs and other QGANs, we have demonstrated the advantages of the quantum-classical hybrid generative model QC-GAN. Compared to classical GANs, experimental results show the powerful quantum capabilities, which can significantly reduce the number of training parameters and simplify the model depth while achieving high-quality generation tasks. Compared to other QGANs, effective integration with classical circuits can achieve better results with limited quantum resources. However, due to linearity property of quantum circuits, the experiment shows a lack of diversity in generating samples, which poses a challenge for generating complex distributions. In addition, due to the current limitations of quantum resources, measurement inaccuracies and high costs, the hardware implementation of the experiment remains a major challenge.
     
\section{CONCLUSION}
This paper utilizes a new hybrid quantum-classical architecture called QC-GAN, which combines the powerful quantum capabilities with the advantages of classical algorithms to better perform generative tasks. Simulation experiments comparing the performance of QC-GAN with traditional GANs in generating tasks show that QC-GAN exhibits good generative ability with a limited number of parameters. Furthermore, QC-GAN is compared with other QGANs and demonstrated better generative ability. Due to the limitations of current quantum resources, many quantum algorithms have their limitations, but the quantum-classical hybrid structure has important significance in the NISQ era. This new structure can provide new methods for quantum computing applications in the near future, as well as new ideas and directions for the development of quantum algorithms. In the future, the quantum-classical hybrid structure has the potential to become an important direction in the field of quantum computing, providing support and guidance for building more powerful and efficient quantum computing applications. Our code is available at \url{https://gitee.com/mindspore/mindquantum/tree/research/paper_with_code}.

\section{ACKNOWLEDGEMENTS}
This work was supported by the National Key Research and Development Program of China under Grant 2022YFB3103100, and the National Natural Science Foundation of China under Grant 62273154.


\begin{thebibliography}{99}
	\bibitem{preskill2012quantum}
	Preskill J. Quantum computing and the entanglement frontier. 2012. 	ArXiv:1203.5813

	\bibitem{biamonte2017quantum}
	Biamonte J, Wittek P, Pancotti N, et al. Quantum machine learning. 	Nature, 2017, 549:195-202
	
	\bibitem{lamata2020quantum}
	Lamata L. Quantum machine learning and quantum biomimetics: A 	perspective. Machine Learning: Science and Technology, 2020, 1: 033002
	
	\bibitem{cerezo2022challenges}
	Cerezo M, Verdon G, Huang H Y, et al. Challenges and opportunities in quantum machine learning. Nature Computational Science, 2022,  2: 567-576
	
	\bibitem{cong2019quantum}
	Cong I, Choi S, and Lukin M D. Quantum convolutional neural networks. Nature Physics, 2019, 15: 1273-1278
	
	\bibitem{blank2020quantum}
	Blank C, Park D K, Rhee J K K, et al. Quantum classifier with tailored quantum kernel. Quantum Information, 2020, 6: 41
	
	\bibitem{lloyd2018quantum}
	Lloyd S and Weedbrook C. Quantum generative adversarial learning. Physical Review Letters, 2018, 121: 040502
	
	\bibitem{dallaire2018quantum}
	Dallaire-Demers P L, Killoran N. Quantum generative adversarial networks. Physical Review A, 2018, 98: 012324
	
	\bibitem{goodfellow2014generative}
	Goodfellow I, Pouget-Abadie J, Mirza M, et al. Generative adversarial nets. Advances in neural information processing systems, 2014. 2672-2680
	
	\bibitem{benedetti2019adversarial}
	Benedetti M, Grant E, Wossnig L, et al. Adversarial quantum circuit learning for pure state approximation. New Journal of Physics, 2019, 21: 043023
	
	\bibitem{zeng2019learning}
	Zeng J, Wu Y, Liu J G, et al. Learning and inference on generative adversarial quantum circuits. Physical Review A, 2019, 99: 052306
	
	\bibitem{zoufal2019quantum}
	Zoufal C, Lucchi A, Woerner S. Quantum generative adversarial networks for learning and loading random distributions. npj Quantum Information, 2019, 5: 103
	
	\bibitem{ahmed2021quantum}
	Ahmed S, Mu{\~n}oz C S, Nori F, et al. Quantum state tomography with conditional generative adversarial networks. Physical Review Letters, 2021, 127: 140502
	
	\bibitem{niu2022entangling}
	Niu M Y, Zlokapa A, Broughton M, et al. Entangling quantum generative adversarial networks. Physical Review Letters, 2022, 128: 220505
	
	\bibitem{li2021quantum}
	Li J, Topaloglu R O, Ghosh S. Quantum generative models for small molecule drug discovery. IEEE Transactions on Quantum Engineering, 2021, 	2: 3103308
	
	\bibitem{nakaji2021quantum}
	Nakaji K, Yamamoto N. Quantum semi-supervised generative adversarial network for enhanced data classification. Scientific reports, 2021, 11: 19649
	
	\bibitem{herr2021anomaly}
	Herr D, Obert B, Rosenkranz M. Anomaly detection with variational
	  quantum generative adversarial networks. Quantum Science and
	  Technology, 2021, 6: 045004
	
	\bibitem{huang2021quantum}
	Huang K, Wang Z A, Song C, et al. Quantum generative adversarial networks with
	  multiple superconducting qubits. npj Quantum Information, 2021, 7: 165
	
	\bibitem{hu2019quantum}
	Hu L, Wu S H, Cai W, et al. Quantum generative adversarial learning in a 	superconducting quantum circuit. Science advances, 2019, 5: 2761
	
	\bibitem{preskill2018quantum}
	Preskill J. Quantum computing in the NISQ era and beyond. Quantum, 2018, 2: 		79.
	
	\bibitem{huang2021experimental}
	Huang H L, Du Y, Gong M, et al. Experimental quantum generative adversarial 	networks for image generation. Physical Review Applied, 2021, 16: 024051
	
	\bibitem{stein2021qugan}
	Stein S A, Baheri B, Chen D, et al. Qugan: A quantum state fidelity based 		generative adversarial network. In: 2021 IEEE International Conference on 		Quantum Computing and Engineering (QCE), 2021. 71-81.
	
	\bibitem{benedetti2019parameterized}
	Benedetti M, Lloyd E, Sack S, et al. Parameterized quantum circuits as 	machine learning models. Quantum Science and Technology, 2019, 4: 	043001
	
	\bibitem{kingma2013auto}
	Kingma D P, Welling M. Auto-encoding variational bayes. 2013. ArXiv: 1312.6114

	\bibitem{ho2020denoising}
	Ho J, Jain A, Abbeel P. Denoising diffusion probabilistic models. Advances in Neural Information Processing Systems, 2020, 33: 6840-6851
	
	\bibitem{duvenaud2015advances}
	Duvenaud D K, Maclaurin D, Iparraguirre J, et al. Convolutional networks on graphs for learning molecular fingerprints. Advances in neural information processing systems, 2015, 28:2224-2232
	
	\bibitem{wang2018esrgan}
	Wang X, Yu K, Wu S, et al. Esrgan: Enhanced super-resolution generative adversarial networks. In: Proceedings of the European Conference on Computer Vision (ECCV). 2018
	
	\bibitem{bai2018sod}
	Bai Y, Zhang Y, Ding M, et al. Sod-mtgan: Small object detection via multi-task generative adversarial network. In: Proceedings of the European Conference on Computer Vision (ECCV). 2018: 206-221.
	
	\bibitem{schlegl2017unsupervised}
	Schlegl T, Seeb{\"o}ck P, Waldstein S M, et al. Unsupervised anomaly detection with generative adversarial networks to guide marker discovery.In: Information Processing in Medical Imaging: 25th International Conference, Boone, 2017. 146-157.
	
	\bibitem{plesch2011quantum}
	Plesch M, Brukner {\v{C}}. Quantum-state preparation with universal gate
	  decompositions. Physical Review A, 2011, 83: 032302
	
	\bibitem{nakaji2022approximate}
	Nakaji K, Uno S, Suzuki Y, et al. Approximate amplitude encoding in shallow parameterized quantum circuits and its application to financial market indicators. Physical Review Research, 2022, 4: 023136.
	
	\bibitem{liu2018differentiable}
	Liu J G, Wang L. Differentiable learning of quantum circuit born
	  machines. Physical Review A, 2018, 98: 062324
	
	\bibitem{Seitzer2020FID}
	Seitzer M. pytorch-fid: FID Score for PyTorch. https://github.com/mseitzer/pytorch-fid, 2020
	
	\bibitem{heusel2017gans}
	Heusel M, Ramsauer H, Unterthiner T, et al. Gans trained by a two time-scale update rule converge to a local nash equilibrium. Advances in neural information processing systems, 2017, 30.
	
	\bibitem{lecun1998gradient}
	LeCun Y, Bottou L, Bengio Y, et al. Gradient-based learning applied to document recognition. Proceedings of the IEEE, 1998, 86: 2278-2324.

	\bibitem{radford2015unsupervised}
	Radford A, Metz L and Chintala S. Unsupervised representation learning with deep convolutional generative adversarial networks. 2015. ArXiv:1511.06434.
	


\end{thebibliography}
\end{document}